# Canonically Relativistic Quantum Mechanics:

# Representations of the Unitary Semidirect Heisenberg Group, $U(1,3) \otimes_s H(1,3)$

Stephen G. Low

Tandem Computers,

Internet: slow@isd.tandem.com

14231 Tandem Blvd, Austin, TX, 78728, USA

Born proposed a unification of special relativity and quantum mechanics that placed position, time, energy and momentum on equal footing through a reciprocity principle and extended the usual position-time and energy-momentum line elements to this space by combining them through a new fundamental constant. Requiring also invariance of the symplectic metric yields $U(1,3)$ as the invariance group, the inhomogeneous counterpart of which is the canonically relativistic group $CR(1,3) = U(1,3) \otimes_s H(1,3)$ where $H(1,3)$ is the Heisenberg Group in 4 dimensions. This is the counterpart in this theory of the Poincaré group and reduces in the appropriate limit to the expected special relativity and classical Hamiltonian mechanics transformation equations. This group has the Poincaré group as a subgroup and is intrinsically quantum with the Position, Time, Energy and Momentum operators satisfying the Heisenberg algebra. The representations of the algebra are studied and Casimir invariants are computed. Like the Poincaré group, it has a little group for a (*"massive"*) rest frame and a null frame. The former is $U(3)$ which clearly contains $SU(3)$ and the latter is $Os(2)$ which contains $SU(2) \otimes U(1)$.

## I    Introduction

Representations of the Poincaré group [1] define fundamental equations of physics associated with spin 0, 1/2, 1 and 3/2: the Klein-Gordon equation, Dirac's equation, Maxwell's equations and the Rariter-Schwinger equation. The representations of the Poincaré group do not, however, give insight into the following key issues:

1. The Heisenberg commutation relations do not appear in the Poincaré algebra and so must be *added on* to the theory. These relations are invariant under the canonical or transformation theory of Dirac [2] that is the quantum generalization of the canonical or symplectic transformations of the classical Hamiltonian theory.

2. The Poincaré group does not encompass the $SU(3)$ symmetries of the strong interactions nor the $SU(2) \times U(1)$ of the weak interactions. These symmetries must also be *added on* to the theory. This has lead to a plethora of higher dimensional theories that attempt to integrate these symmetries with the underlying position-time space transformations defined by the Poincaré group. These theories must explain why the higher dimensions are not directly observable.

3. While the fundamental physical concepts of mass and spin (or helicity) may be ascribed to the interpretation of the two Casimir invariants of the Poincaré group in the rest (or zero mass) frame, the theory gives no insight into the discrete spectra of the particle rest mass invariants.

4. Many of the theories based on the Poincaré representations are plagued with infinities that must be removed through renormalization theories. The physical foundation of this renormalization has its roots in an invariant *cut-off* in the distance and time dimension [3]. In many cases, an equivalent way of describing this is to say that the acceleration or rate of change of momentum of the particles is bounded [4][5][6].

To address the first of these issues the theory must have as invariants both the quantum canonical commutation relations:

$$[P_i, Q_j] = ihI\delta_{ij}, \quad [E, T] = -ihI , \qquad (1)$$

and the invariant mass and line elements of the Lorentz group of special relativity

$$\tau^2 = T^2 - \delta^{ij} Q_i Q_j / c^2 \qquad m^2 = E^2/c^2 - \delta^{ij} P_i P_j . \qquad (2)$$

$\{T, E, Q_i, P_i\}$ are the time, energy, position and momentum degrees of freedom and $i, j, \ldots = 1, 2, 3$. Born [4], Caianiello [5] and the author [6][7] have argued that for such a theory, the line elements in (2) must be combined into a single line element through the introduction of a new fundamental physical constant. In Born's theory, the constant has the dimension and interpretation of a minimum length, $a$, in Caianiello's theory of maximum acceleration $A_{max}$ and in the author's, a universal upper bound on rate of change of momentum $b$:

$$s^2 = \tau^2 + m^2/b^2 = T^2 - \delta^{ij} Q_i Q_j / c^2 - \delta^{ij} P_i P_j / b^2 + E^2/(bc)^2 \qquad (3)$$

Born's minimum distance $a$ may be then defined as $a = \sqrt{hc/b}$ and Caianiello's maximal acceleration $A_{max}$ as

$$A_{max} = \frac{hc}{m_0 a^2} = \frac{b}{m_0}. \tag{4}$$

In this equation, $m_0$ is some fundamental mass. In fact, as there are only three dimensionally independent universal constants, the measures of time, distance, momentum and energy may be given in terms of the constants appearing in equations (1) and (3):

$$\lambda_t = \sqrt{h/bc}, \lambda_q = \sqrt{hc/b}, \lambda_p = \sqrt{hb/c}, \lambda_e = \sqrt{hcb}. \tag{5}$$

All other physical constants may then be expressed as dimensionless multiples of expressions constructed from these measures. For example, the gravitational constant $G$ and the electric charge $e$ are given by

$$G = \alpha_G c^4/b, \qquad e^2 = \alpha_e hc. \tag{6}$$

$\alpha_e$ is the dimensionless fine structure and if $\alpha_G = 1$, the measures given in (5) are the usual Planck measures.

With the definition of the line element in equation (3), the concept of an absolute rate of change of momentum, or force, loses its meaning as did absolute velocity in the special relativistic case. Rather, forces are only defined between bodies and their associated frames of reference. Further more, the simple Euclidean summation of forces is replaced with a relativistic formula analogous to the special relativistic velocity addition equation that ensures that the rates of change of momentum are bounded by the universal constant $b$. This is studied in [6] and summarized in Section II .

The definition of $b$ given by (3) enables a fundamental geometric or group theoretic interpretation of these fundamental dimensional invariants analogous to the generalization from the Euclidean group of Newton's mechanics to the Lorentz group of Einstein's mechanics. Equation (1) is invariant under the symplectic group $Sp(8)$ and equation (3) is invariant under the orthogonal group $O(2, 6)$. The intersection of these groups, which leaves both invariants invariant is the 16 generator $U(1, 3)$ group. This may be more clearly seen in what follows.

$$X_a = \{T/\lambda_t, Q_i/\lambda_q\}, \qquad Y_a = \{E/\lambda_e, P_i/\lambda_p\} \tag{7}$$

where $a, b, \ldots = 0, 1, 2, 3, 4$. In terms of this basis, the orthogonal metric of $O(2, 6)$ and the symplectic metric of $Sp(8)$ may then be written as

$$s^2 = \lambda_t^2 \eta^{ab}(X_a X_b + Y_a Y_b)$$
$$\sigma = \eta^{ab}(Y_a X_b - X_b Y_a) \tag{8}$$

where $\eta^{ab} = diag\{-1, 1, 1, 1\}$. Defining the usual complex basis

$$A_a^+ = (X_a + iY_a)/\sqrt{2}, \qquad A_a^- = (X_a - iY_a)/\sqrt{2} \tag{9}$$

The Hermitian $U(1, 3)$ metric is:

$$\zeta^2 = \eta^{ab} A_a^+ A_b^-, \tag{10}$$

Expanding shows that $\zeta^2 = \left(s^2/\lambda_t^2 + \sigma\right)/2$. It is important to note that we could have equally well defined

$$\tilde{X}_a = \{T/\lambda_t, -P_i/\lambda_p\}, \tilde{Y}_a = \{E/\lambda_e, Q_i/\lambda_q\},$$
$$\tilde{A}_a = \tilde{X}_a \ i\tilde{Y}_a \tag{11}$$
$$\zeta^2 = \eta^{ab}\tilde{X}_a\tilde{X}_b.$$

This tilde transform $\{T, E, Q_i, P_i\} \to \{T, E, -P_i, Q_i\}$ is precisely Born's reciprocity transformation [4].

Equation (10) is the defining invariant of the group $U(1, 3)$. The author shows in [6] that this equation contracts to the expected Lorentz and Euclidean equations in the limits $b \to \infty$ and $b, c \to \infty$. Again, this is summarized in Section II.

Before turning to this, it should be clear that $U(1, 3)$ is the homogeneous transformation group in the theory we are developing in the same way that the Lorentz group $SO(1, 3)$ is the homogeneous group of the Poincaré group $SO(1, 3) \otimes_s T(4)$ of special relativity. The natural inhomogeneous group for the current theory is then the Canonical Relativistic group:

$$CR(1, 3) = U(1, 3) \otimes_s H(1, 3). \tag{12}$$

$H(1, 3)$ is the 9 generator Heisenberg Group with the corresponding Lie Algebra defined by (1).

The name Canonical Relativistic group is to emphasize that it has, as invariants, both the canonical or symplectic structure of quantum mechanics given in (1) and the relativistic line elements (2) combined in (3). Its algebra has 25 generators compared to the 10 of the Poincaré algebra.

The generators of the Inhomogeneous group $H(1, 3)$ do not all commute and so cannot be simultaneously diagonalized like the commuting generators of $T(4)$ in the Poincaré group. Consequently the space $H(1, 3) \approx CR(1, 3)/U(1, 3)$ is intrinsically quantum. Four generators in the set of eight, $\{T, E, Q_i, P_i\}$, may be simultaneously diagonalized. ($I$ commutes with all generators and so may be always diagonalized.) In fact, if the degrees of the quantum phase space are arranged in the square as illustrated in Figure 1, then only the degrees of freedom on one side of the figure at a time may be simultaneously diagonalized.

Figure 1    Illustration of the quantum phase space.

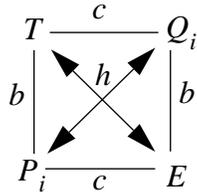

The physical constants show the dimensional relationships. The homogeneous $U(1, 3)$ group transforms all degrees of freedom into each other and so the concepts of independent position-time space and energy-momentum spaces disappears. These fundamental physical degrees of freedom may be transformed into each other under the general $U(1, 3)$ transformations. However, these effects are apparent only when rates of change of position and momentum approach $c$ and $b$ respectively. Thus if these constants are very large, the general effects are fully visible only for very high velocity, very strongly interacting particles.

In the following section, the results obtained in [6] that show the behavior of this group in the limits of $c$ and $b$ becoming infinite are briefly summarized. In the non-relativistic limit where both the rate of change of position and momentum are small relative to their respective constants $c$ and $b$ the space *breaks* into the independent time, position, energy and momentum degrees of freedom in the same sense that the Lorentz relativistic position-time space of Einstein's mechanics *breaks* into the Euclidean position and time space of Newton's mechanics.

The remaining sections introduce the representation theory of the canonical relativistic group $CR(1, 3) = U(1, 3) \otimes_s H(1, 3)$. Returning to the points raised in the introduction regarding limitations in the Poincaré group, note that the canonical relativistic group has the following properties:

1. The Heisenberg commutation relations are intrinsically part of the algebra.

2. $SU(3) \otimes U(1)$ is the *Little Group* (rest frame) for the invariant $\zeta^2 > 0$ and $(SU(2) \otimes U(1)) \otimes_S H(2)$ for the case $\zeta^2 = 0$. This corresponds to the Poincaré Little Groups $SO(3)$ and $SO(2) \otimes_S T(2)$ for the cases $m^2 > 0$ and $m^2 = 0$ respectively.

3. As will be shown in Section III, all spectra are discrete.

4. This was Born's basic premise for the reciprocity principle. From the comments that a bound in rate of change of momentum is in some sense equivalent to the cut-off, there is some hope in this direction. Definitive determination requires the development of a full theory towards which the ideas presented here are a small step.

## II  $CR(1, 3)$ Algebra and Transformation Equations

The $CR(1, 3)$ algebra is given by

$$[Z_{ab}, Z_{cd}] = \eta_{bc} Z_{ad} - \eta_{ad} Z_{cb}$$
$$\left[Z_{ab}, A_c^+\right] = -\eta_{ac} A_b^+, \quad \left[Z_{ab}, A_c^-\right] = \eta_{bc} A_a^- \tag{13}$$
$$\left[A_a^+, A_b^-\right] = \eta_{ab} I$$

where $a, b, \ldots = 0, 1, 2, 3$. The $Z_{ab}$ generators span the $U(1, 3)$ subalgebra and the generators $\{A_a, I\}$ span the Heisenberg subalgebra $H(1, 3)$. Note that

$$A_a^+ = A_a^{-\dagger}$$
$$Z_{ab} = Z_{ba}^{\dagger} \tag{14}$$

The $U(1, 3)$ transformation equations are then given by the exponential of the adjoint representation. A general element of the $U(1, 3)$ algebra is

$$Z = \psi^{ab} Z_{ab} \tag{15}$$

The exponential of the adjoint action is given by

$$A'_d = \Omega^c_d\left(\psi^{ab}\right) A_c = e^{[Z,\ ]} A_d$$
$$= A_d + [Z, A_d] + \tfrac{1}{2!}[Z, [Z, A_d]] + \dots \quad (16)$$

The infinitesimal equations, valid for sufficiently small $\psi^{ab}$ are given by the first two terms in the expansion. Define

$$Z_{ab} = \tfrac{1}{2}(M_{ab} + iL_{ab})$$
$$\psi^{ab} = \varphi^{ab} - i\phi^{ab} \quad (17)$$

where

$$M_{ab} = M_{ba} \quad L_{ab} = -L_{ba} \quad \varphi^{ab} = \varphi^{ba} \quad \phi^{ab} = -\phi^{ba} \quad (18)$$

and are real. The general element of the $U(1, 3)$ algebra is then

$$Z = \psi^{ab} Z_{ab} = \phi^{ab} L_{ab} + \varphi^{ab} M_{ab} \quad (19)$$

Recalling the definition (9), $A_a = X_a\ iY_a$, the commutation relations for the algebra of $CR(1, 3)$ in terms of the 25 real generators $\{L_{ab}, M_{ab}, X_a, Y_a, I\}$ is given by:

$$[L_{ab}, L_{cd}] = \eta_{ad} L_{bc} + \eta_{bc} L_{ad} - \eta_{ac} L_{bd} - \eta_{bd} L_{ac}$$
$$[L_{ab}, M_{cd}] = -\eta_{ad} M_{bc} + \eta_{bc} M_{ad} - \eta_{ac} M_{bd} + \eta_{bd} M_{ac}$$
$$[M_{ab}, M_{cd}] = -\eta_{ad} L_{bc} - \eta_{bc} L_{ad} - \eta_{ac} L_{bd} - \eta_{bd} L_{ac} \quad (20)$$
$$[Y_a, X_b] = i\eta_{ab} I$$

$$[L_{ab}, Y_c] = i(\eta_{ac} Y_b - \eta_{bc} Y_a) \quad [L_{ab}, X_c] = i(\eta_{ac} X_b - \eta_{bc} X_a)$$
$$[M_{ab}, Y_c] = i(\eta_{ac} X_b + \eta_{bc} X_a) \quad [M_{ab}, X_c] = -i(\eta_{ac} Y_b + \eta_{bc} Y_a)$$

It is clear that the $L_{ab}$ are the generators of the $SO(1, 3)$ algebra and that $\{L_{ab}, Y_a\}$ are the generators of the usual Poincaré algebra. It is also interesting to note that for the tilde basis given in equation (11), that there are an identical set of relations results with corresponding tilde generators $\{\tilde{L}_{ab}, \tilde{M}_{ab}, \tilde{X}_a, \tilde{Y}_a, I\}$. However in this case, the Lorentz transformations of the $\{\tilde{L}_{ab}\}$ are in the time-momentum and energy-position degrees of the quantum phase space illustrated in Figure 1 instead of the usual time-position and energy-momentum degrees of freedom.

To understand this more clearly and to enable the limiting behavior of the algebra to be more clearly understood, it is convenient to expand the definition further. Define

$$\beta^i = c\phi^{0i}, \quad \gamma^i = b\phi^{0i}, \quad \alpha^i = \varepsilon^i_{jk}\phi^{jk}, \quad \vartheta = bc\phi^{00}, \quad \theta^{ij} = \phi^{ij}/bc \qquad (21)$$

and

$$K_i = -L_{0i}, \quad N_i = M_{0i}, \quad J_i = -\varepsilon_i^{jk}L_{jk}, \quad R_0 = M_{00} \qquad (22)$$

Using the definitions (7), the transformation equations are then

$$\begin{aligned}
T' &= T + \beta^i Q^i/c^2 + \gamma^i P^i/b^2 + \vartheta E/c^2 b^2 \\
E' &= E - \gamma^i Q^i + \beta^i P^i - \vartheta T \\
Q'_i &= Q_i + \varepsilon^k_{ij}\alpha^j P_k + \beta^i T - \gamma^i E/b^2 - \theta^{ij} P_j/b^2 \\
P'_i &= P_i + \varepsilon^k_{ij}\alpha^j P_k + \beta^i E/c^2 - \gamma^i T + \theta^{ij} Q_j/c^2
\end{aligned} \qquad (23)$$

From these equations, it is clear that the $J_i$ are the usual rotation generators and $K_i$ the usual Lorentz boost generators for position-time or momentum-energy hyperbolic rotations. The new generators $N_i$ are corresponding Lorentz boost generators for momentum-time or position-energy hyperbolic rotations.

The limit of $b, c \to \infty$ may be straightforwardly determined from (23):

$$\begin{aligned}
T' &= T \\
E' &= E - \gamma^i Q^i + \beta^i P^i - \vartheta T \\
Q'_i &= Q_i + \varepsilon^k_{ij}\alpha^j Q_k + \beta^i T \\
P'_i &= P_i + \varepsilon^k_{ij}\alpha^j P_k - \gamma^i T
\end{aligned} \qquad (24)$$

and it follows that through the simple Wigner-Inönü (commuting) contractions that:

$$\lim_{b, c \to \infty} \{J_i, K_i, N_i, R, M_{ij}\} = \{J_i, G_i, F_i, R_0, 0\} \qquad (25)$$

satisfy the commutation relations

$$[J_i, J_j] = \varepsilon^k_{ij}J_k, \quad [J_i, G_j] = \varepsilon^k_{ij}G_k, \quad [J_i, F_j] = \varepsilon^k_{ij}F_k,$$

$$[F_i, G_j] = \delta_{ij}R_0, \quad [J_i, Q_j] = i\varepsilon^k_{ij}Q_k, \quad [J_i, P_j] = i\varepsilon^k_{ij}P_k, \qquad (26)$$

$$[G_i, Q_j] = i\delta_{ij}T, \quad [F_i, P_j] = -i\delta_{ij}T, \quad [G_i, E] = iP_i, \quad [F_i, E] = -iP_i,$$

$$[R_0, E] = -iT, \quad [P_i, Q_j] = ih\delta_{ij}I, \quad [E, T] = -ihI$$

As expected, in the Newtonian limit the generators $G_i$ and $F_i$ give simple Euclidean addition for velocity $\beta^2 \ll c^2$ and force $\gamma^2 \ll b^2$.

The partial contraction $b \to \infty$ results in the usual Lorentz transformations on position-time and energy-momentum spaces and with the $M_{ab}$ commuting with themselves and behaving as a symmetric (2,0) tensor under Lorentz transformations. The limit $c \to \infty$ results in identical results with the tilde generators: Lorentz transformations on the momentum-time and energy position spaces with the $M_{ab}$ commuting with themselves and behaving as a symmetric (2,0) tensor under the tilde Lorentz transformations

Returning now to the non-contracted algebra, the finite transformations may be computed using equation (16). The full transformations may be decomposed into $U(3)$ rotations and the velocity and force boosts. Taking the boost only case,

$$A'_a = \Omega(\beta^i, \gamma^i)_a^b A_b = e^{[\beta^i K_i + \gamma^i N_i,\,]} A_a. \qquad (27)$$

The corresponding transformation equations are

$$(28)$$

$$T' = \cosh\rho\, T + \frac{\sinh\rho}{\rho}\left(\beta^i Q_i/c^2 + \gamma^i P_i/b^2\right)$$

$$E' = \cosh\rho\, E + \frac{\sinh\rho}{\rho}\left(-\gamma^i Q_i + \beta^i P_i\right)$$

$$Q'_i = Q_i + \left(\frac{\cosh\rho - 1}{\rho^2}\right)\left(\beta^i\beta^j/c^2 + \gamma^i\gamma^j/b^2\right)Q_j + \frac{\sinh(\rho)}{\rho}\left(\beta^i T - \gamma^i E/b^2\right)$$

$$P'_i = P_i + \left(\frac{\cosh\rho - 1}{\rho^2}\right)\left(\beta^i\beta^j/c^2 + \gamma^i\gamma^j/b^2\right)P_j + \frac{\sinh(\rho)}{\rho}\left(\beta^i E/c^2 + \gamma^i T\right)$$

where

$$\rho = \sqrt{\delta_{ij}\left(\beta^i\beta^j/c^2 + \gamma^i\gamma^j/b^2\right)}. \qquad (29)$$

Note again that, in the limit $b \to \infty$, these equations reduce to the expected special relativity equations. Furthermore, the limit $c \to \infty$, applicable to the case where relative velocities are small but relative forces are not, results in a similar set of Lorentz transformations. The addition equation for the relative rates of change of position and momentum that may be derived straightforwardly from (28) shows that $c$ and $b$ are the respective upper bounds of their magnitudes.

**III  Representations of $CR(1, 3)$**

The unitary representations of the Canonical Relativistic group may be developed using Wigner's *Little Group* approach [1][8][9] that is generally also used for the study of the Poincaré group. $CR(1, 3)$ is rank 5 and consequently there are 5 Casimir invariants that may be used to label the representation. Define

$$W_{ab} = A^-_a A^+_b - I Z_{ab} \qquad (30)$$

and a straightforward computation yields

$$[A_a, W_{bc}] = 0$$
$$[Z_{ab}, W_{cd}] = \eta_{bc} W_{ad} - \eta_{ad} W_{cb} \, , \qquad (31)$$

from which it follows that the Casimir invariant are

$$C_0(CR(1,3)) = I \qquad (32)$$
$$C_1(CR(1,3)) = \eta^{ab} W_{ab}$$
$$C_2(CR(1,3)) = \eta^{ad} \eta^{bc} W_{ab} W_{cd}$$
$$C_3(CR(1,3)) = \eta^{af} \eta^{bc} \eta^{de} W_{ab} W_{cd} W_{ef}$$
$$C_4(CR(1,3)) = \eta^{ah} \eta^{bc} \eta^{de} \eta^{fg} W_{ab} W_{cd} W_{ef} W_{gh}$$

Expanding out the $C_1$ Casimir invariant yields

$$C_1(CR(1,3)) = -T^2 - \frac{E^2}{(cb)^2} + \frac{Q^2}{c^2} + \frac{P^2}{b^2} - \delta^{ij}[P_i, Q_i] + [E, T] + Z^\circ \qquad (33)$$

where $Z^\circ = \delta^{ab} Z_{ab}$ is the $U(1)$ generator in the decomposition

$$U(1, 3) = U(1) \otimes SU(1, 3) .\qquad(34)$$

The Little Group is the subgroup of the full group that leaves invariant a standard vector. The space divides, as in the Poincaré case into null, time-like and space-like cases. In this case, the time-like case is for the physical regions of the quantum phase space where the bounds $\beta^2 < c^2$ and $\gamma^2 < b^2$. Note as the generators $K_i$ and $N_i$ do not commute, these generators may not be simultaneously diagonalized and hence only $\beta$ or $\gamma$ is precisely observable at a given time. In the null case, either $\beta^2 = c^2$ or $\gamma^2 = b^2$ holds. Feynman's picture of the electron comes to mind where the electron zig-zagging with an instantaneous velocity of $c$ even though its average relative velocity is less and in fact, could be zero [10]. The corresponding momentum picture under this theory would have the rate of change of momentum exhibiting the same zig-zag with an instantaneous value of $b$ if it is to be described by the null picture. The non-physical regions are when the bounds are exceeded and are not considered further here.

For the time-like case, the standard vector $a_0$ may be defined as $a_0 = (1, 0, 0, 0)$ and for the null case, as $a_0 = (1, 0, 0, 1)$.

### 3.1 Time Like Case

In the time-like case, consider the subgroup $CR(1, 3) \supset U(3) \otimes Os(1)$ where $U(3)$ is the Little group with generators $\{Z_{ij}\}$ and $Os(1)$ is the standard vector group spanned by the generators $\{E, T, I, Z^\circ\}$ that define an Oscillator Group algebra [11]. (In the Poincaré case, the standard vector group is the trivial $T(1)$. The Little Group may be decomposed into $U(3) = U(1) \otimes SU(3)$.

Define the standard $SU(3)$ generators by

$$\begin{aligned} U^+ &= Z_{31}, & V^+ &= Z_{12}, & T^+ &= Z_{23}, & T^\circ &= (-Z_{11} + Z_{22})/2, \\ U^- &= Z_{13}, & V^- &= Z_{21}, & T^- &= Z_{32}, & Y &= (Z_{22} + Z_{33} - 2Z_{11})/3 \end{aligned}\qquad(35)$$

Also, define the $Os(1)$ generator $R$ and the $U(1)$ generator $Z^\circ$ by

$$R = \eta^{ab} Z_{ab}, \qquad Z^\circ = \delta^{ij} Z_{ij}\qquad(36)$$

The general matrix elements for the generators of $U^{\mu_1, \mu_2}(SU(3))$ have been computed by Lurié and MacFarlane [12]. The general expressions result in considerable algebraic complexity when computing the full $CR(1, 3)$ representations. However, if the subset of representations with $\mu_2 = 0$ is considered, a dramatic simplification results. As the intent of this paper is to clearly demonstrate the concepts behind the canonical relativistic group, we choose to restrict the discussion to the simplified case and present the fully general rep-resentations in a subsequent paper.

The action of the unitary representations $U^{\mu_1, 0}(SU(3))$ of the $SU(3)$ group on a basis may be written as

$$\begin{aligned}
T^+|m,j\rangle &= \sqrt{((j+\kappa_1)/2 + m + 1)((j+\kappa_1)/2 - m)}\,|m+1,j\rangle \\
T^-|m,j\rangle &= \sqrt{((j+\kappa_1)/2 + m)((j+\kappa_1)/2 - m + 1)}\,|m-1,j\rangle \\
U^+|m,j\rangle &= \sqrt{(\nu_1 - j)((j+\kappa_1)/2 - m + 1)}\,|m-\tfrac{1}{2},j+1\rangle \\
U^-|m,j\rangle &= \sqrt{(\nu_1 - j + 1)((j+\kappa_1)/2 - m)}\,|m+\tfrac{1}{2},j-1\rangle \\
V^+|m,j\rangle &= \sqrt{(\nu_1 - j + 1)((j+\kappa_1)/2 + m)}\,|m-\tfrac{1}{2},j-1\rangle \\
V^-|m,j\rangle &= \sqrt{(\nu_1 - j)((j+\kappa_1)/2 + m + 1)}\,|m+\tfrac{1}{2},j+1\rangle \\
Y|m,j\rangle &= \left(j - \tfrac{1}{3}(2\nu_1 - \kappa_1)\right)|m,j\rangle \\
T^\circ|m,j\rangle &= m|m,j\rangle
\end{aligned} \tag{37}$$

It may be verified that this satisfies the $SU(3)$ algebra most straightforwardly with a symbolic computation package such as Mathematica [13].

$\{j, \nu_1, \kappa_1\}$ are integer and $m$ is half integer. A direct computation shows that this has the expected Casimir invariants with $\mu_1 = \nu_1 + \kappa_1$:

$$\begin{aligned}
C_2(SU(3)) &= 2\mu_1(\mu_1 + 3)/3 \\
C_3(SU(3)) &= \mu_1(\mu_1 + 3)(2\mu_1 - 6)/9
\end{aligned} \tag{38}$$

The $U(1)$ representation is trivially

$$Z^\circ|n\rangle = (n + \kappa_2)|n\rangle \tag{39}$$

In this representation, $\{n, \kappa_2\}$ are integer.

Unitary representations of both the Oscillator group are given by [11]:

$$\begin{aligned}
A^+|k\rangle &= \sqrt{\kappa_0(k+1)}\,|k+1\rangle \\
A^-|k\rangle &= \sqrt{\kappa_0 k}\,|k-1\rangle \\
R|k\rangle &= (k + \nu_2 + 1)|k\rangle
\end{aligned} \tag{40}$$

with $\{k, \nu_2\}$ integer and $\kappa_0$ real and positive. Its Casimir Invariants are

$$I|k\rangle = \kappa_0|k+1\rangle$$
$$\left(A^- A^+ - IR\right)|k\rangle = -\kappa_0 \nu_2|k\rangle \tag{41}$$

The basis for $CR(1, 3)$ may be constructed as the direct product

$$|m, j, n, k\rangle = |m, j\rangle \otimes |n\rangle \otimes |k\rangle. \tag{42}$$

The full representation may be computed by the action of the remaining boost generators on the basis vectors. In order for the algebra to close, the parameters $\{\nu_1, \nu_2\}$ appearing in equations (37) and (40) must satisfy:

$$\nu_1 = n \tag{43}$$

Then, defining the remaining $SU(1, 3)$ generators:

$$B_i = N_i \quad iK_i \tag{44}$$

their action on the basis is given by:

$$\begin{aligned}
B_1^+|m, j, n, k\rangle &= \sqrt{(n-j+1)(k+1)}\,|m, j, n+1, k+1\rangle \\
B_1^-|m, j, n, k\rangle &= \sqrt{(n-j)\,k}\,|m, j, n-1, k-1\rangle \\
B_2^+|m, j, n, k\rangle &= \sqrt{((j+\kappa_1)/2 + m + 1)(k+1)}\,|m+\tfrac{1}{2}, j+1, n+1, k+1\rangle \\
B_2^-|m, j, n, k\rangle &= \sqrt{((j+\kappa_1)/2 + m)\,k}\,|m-\tfrac{1}{2}, j-1, n-1, k-1\rangle \\
B_3^+|m, j, n, k\rangle &= \sqrt{((j+\kappa_1)/2 - m + 1)(k+1)}\,|m-\tfrac{1}{2}, j+1, n+1, k+1\rangle \\
B_3^-|m, j, n, k\rangle &= \sqrt{((j+\kappa_1)/2 - m)\,k}\,|m+\tfrac{1}{2}, j-1, n-1, k-1\rangle
\end{aligned} \tag{45}$$

It follows from the commutation relations that the remaining inhomogeneous generators in this representation are

$$A_1^+ |m, j, n, k\rangle = \sqrt{\kappa_0 (n-j)} \, |m, j, n-1, k\rangle$$
$$A_1^- |m, j, n, k\rangle = \sqrt{\kappa_0 (n-j+1)} \, |m, j, n+1, k\rangle$$
$$A_2^+ |m, j, n, k\rangle = \sqrt{\kappa_0 ((j+\kappa_1)/2 + m)} \, |m-\tfrac{1}{2}, j-1, n-1, k\rangle$$
$$A_2^- |m, j, n, k\rangle = \sqrt{\kappa_0 ((j+\kappa_1)/2 + m + 1)} \, |m+\tfrac{1}{2}, j+1, n+1, k\rangle$$
$$A_3^+ |m, j, n, k\rangle = \sqrt{\kappa_0 ((j+\kappa_1)/2 - m)} \, |m+\tfrac{1}{2}, j-1, n-1, k\rangle$$
$$A_3^- |m, j, n, k\rangle = \sqrt{\kappa_0 ((j+\kappa_1)/2 - m + 1)} \, |m-\tfrac{1}{2}, j+1, n+1, k\rangle$$

(46)

A direct computation then gives the Casimir invariants defined in (32):

$$C_0(CR(1,3)) = \kappa_0$$
$$C_1(CR(1,3)) = \kappa_0 (\kappa_1 - \kappa_2 + \nu_2)$$
$$C_2(CR(1,3)) = \kappa_0^2 \left( \frac{(\kappa_1 - \kappa_2)^2}{3} + \nu_2^2 \right)$$
$$C_3(CR(1,3)) = \kappa_0^3 \left( \frac{(\kappa_1 - \kappa_2)^3}{9} + \nu_2^3 \right)$$

(47)

As the equations (47) have only three independent parameters, this is not the most general representation of the rank 5 group requiring 5 parameters. It demonstrates the problem is tractable and should be directly applicable in a restricted physical case.

### 3.2 Null Case

In the null case, consider the subgroup $CR((1,3) \supset Os(2) \otimes CE(1))$ where $Os(2) = U(2) \otimes_s H(2)$ is the Little Group and $CE(1) = U(1) \otimes_s T(2)$ is the complex euclidean standard vector group. The generators of the Little Group are

$$C_i^+ = Z_{i0} - Z_{i3}, \qquad C_i^- = Z_{0i} - Z_{3i}, \qquad I^\circ = Z_{00} - Z_{30} + Z_{33} - Z_{03} \qquad (48)$$

where in this section $i, j, \ldots = 1, 2$. It is clear by construction that the standard vector $a^\circ = (1, 0, 0, 1)$ is left invariant by the action of this group. The commutation relations are the $Os(2)$ algebra:

$$[Z_{ij}, Z_{kl}] = \delta_{jk} Z_{il} - \delta_{il} Z_{kj}$$
$$[Z_{ij}, C_k^+] = \delta_{jk} C_i^+, \quad [Z_{ij}, C_k^-] = -\delta_{ik} C_j^-  \tag{49}$$
$$[C_i^+, C_j^-] = -\delta_{ij} I°$$

On the other hand, the standard vector group has the generators

$$A° = (A_3 \pm A_0)/2$$
$$R = -Z_{00} + Z_{11} + Z_{22} + Z_{33}  \tag{50}$$

which satisfy the $E(2)$ commutation relations:

$$[A°, R] = A°, \qquad [A°^+, A°^-] = 0  \tag{51}$$

It can be verified directly that the generators of the Little Group and the Standard Vector group commute as required.

The remaining non-zero commutation relations are:

$$[Z_{ij}, A_k^+] = -\delta_{ik} A_j^+, \quad [Z_{ij}, A_k^-] = \delta_{jk} A_i^-$$
$$[A_i^+, A_j^-] = \delta_{ij} I$$
$$[C_i, A_j] = 2\delta_{ij} A°  \tag{52}$$
$$[A_i, R] = A_i$$

This is a total of 17 generators: $\{Z_{ij}, C_k, A_k, A°, R, I°, I\}$. The remaining 8 generators in the full $CR(1, 3)$ algebra that do not leave the null vector $a°$ invariant are identically zero. Note that the generators $\{Z_{ij}, A_k\}$ also define an $Os(2)$ algebra.

The $Os(2)$ representations may be computed following the arguments of the previous section. Defining:

$$J^+ = Z_{12}, \qquad J^- = Z_{21}, \qquad J_3 = Z_{11} - Z_{22}, \qquad Z° = Z_{11} + Z_{22}  \tag{53}$$

it follows that the action on the basis vectors are

$$J^+|m, j\rangle = \sqrt{((j + \kappa_1)/2 + m + 1)((j + \kappa_1)/2 - m)}\,|m + 1, j\rangle$$

$$J^-|m, j\rangle = \sqrt{((j + \kappa_1)/2 + m)((j + \kappa_1)/2 - m + 1)}\,|m - 1, j\rangle$$

$$J_3|m, j\rangle = 2m|m, j\rangle$$

$$Z_\circ|m, j\rangle = (j + \kappa_1/2 - \kappa_2)\,|m, j\rangle$$

$$C_1^+|m, j\rangle = \sqrt{\kappa_0((j + \kappa_1)/2 + m + 1)}\,|m + \tfrac{1}{2}, j + 1\rangle \tag{54}$$

$$C_2^+|m, j\rangle = \sqrt{\kappa_0((j + \kappa_1)/2 - m + 1)}\,|m - \tfrac{1}{2}, j + 1\rangle$$

$$C_1^-|m, j\rangle = \sqrt{\kappa_0((j + \kappa_1)/2 + m)}\,|m - \tfrac{1}{2}, j - 1\rangle$$

$$C_2^-|m, j\rangle = \sqrt{\kappa_0((j + \kappa_1)/2 - m)}\,|m + \tfrac{1}{2}, j - 1\rangle$$

$$I^\circ|m, j\rangle = \kappa_0|m, j\rangle$$

The standard vector group generator actions are

$$A^{\circ+}|k\rangle = \kappa_3|k - 1\rangle$$

$$A^{\circ-}|k\rangle = \kappa_4|k + 1\rangle \tag{55}$$

$$R|k\rangle = (k + \nu_1)\,|k\rangle$$

Again, the basis of the full group and algebra is the direct product of the basis of the Little Group and the Standard Vector Group: $|m, j, k\rangle = |m, j\rangle \otimes |k\rangle$. The generators in equations (54) and (55) act on this basis in the usual direct product manner. The remaining generator actions are

$$A_1^+|m, j\rangle = \sqrt{\kappa_5((j + \kappa_1)/2 + m)}\,|m - \tfrac{1}{2}, j - 1, k - 1\rangle$$

$$A_2^+|m, j\rangle = \sqrt{\kappa_5((j + \kappa_1)/2 - m)}\,|m + \tfrac{1}{2}, j - 1, k - 1\rangle$$

$$A_1^-|m, j\rangle = \sqrt{\kappa_5((j + \kappa_1)/2 + m + 1)}\,|m + \tfrac{1}{2}, j + 1, k + 1\rangle \tag{56}$$

$$A_2^-|m, j\rangle = \sqrt{\kappa_5((j + \kappa_1)/2 - m + 1)}\,|m - \tfrac{1}{2}, j + 1, k + 1\rangle$$

$$I^\circ|m, j\rangle = \kappa_5|m, j, k\rangle$$

In order for the algebra to close, the constants must satisfy the relations $\kappa_5 = \kappa_0$, $\kappa_4 = \kappa_3 = -\kappa_0/2$, and $\nu_1 = j - k + \kappa_2$.

The three independent null Casimir Invariants are:

$$C_0(CR(1, 3)) = \kappa_0$$
$$C_1(CR(1, 3)) = \kappa_0(\kappa_1 - \kappa_2)$$
$$C_2(CR(1, 3)) = \frac{\kappa_0^2}{4}\left((\kappa_1 - \kappa_2)^2 + 9\kappa_2^2\right)$$

(57)

## IV  Discussion

From the analysis, it is apparent that the Canonical Relativistic group has many of the mathematical properties required to address the four issues highlighted in the introduction. The quantum commutation relations are intrinsically defined in the algebra, both $SU(3)$ and $SU(2) \otimes U(1)$ appear as subgroups of the Little groups for the generalized time-like and null cases. Absolute forces loose their meaning and are defined only between reference frames associated with particle states. These forces are bounded by the universal constant $b$ that provides the third constant to define the dimensional measures. In the limit of small rates of change of position and momentum relative to their bounding constants, the theory reduces to the expected Newtonian limit. Furthermore, the bound in relative forces is qualitatively equivalent to a distance or time cut-off that results in the promise of a finite theory. The spectra of the Casimir invariants are discrete.

The most remarkable physical consequence of the Canonical Relativistic group is that position-time space is no longer invariant or *absolute*. Rather, the full eight degrees of freedom of the quantum phase space with the physical degrees of freedom of time, energy, momentum and position must be considered. This is a remarkable space in that the degrees of freedom cannot all be simultaneously diagonalized, but rather may only be done so in pairs as illustrated with Figure 1. These effects become particularly important as the rates of change of position and momentum (relative velocity and force) approach the bounds $c$ and $b$. Such a condition may be present in the early universe and one can envision a theory where the *rotation* or *condensation* of the large energy and momentum degrees of freedom into the position and time degrees of freedom as the universe cools and consequently expands in the position and time directions through this transformation.

While the mathematics is intriguing, the question is how to make a connection with physical theory. The next step would appear to be to study the reduction of the representations of the full algebra with respect to the usual Poincaré algebra that is a subalgebra (and in particular the Casimir invariants) and then to derive field equations that are the Canonical Relativistic group generalization of the Dirac, Maxwell, Klein-Gordon equations that may be derived from the Poincaré group. These are non trivial tasks and must be the subject of a subsequent paper.  It is hoped that the ideas presented here will stimulate research into this mathematically remarkable class of theories that have the promise to address key out-standing physical questions.